\documentclass[aps,prl,reprint,superscriptaddress]{revtex4-1}

\usepackage{graphicx}
\usepackage{threeparttable}
\usepackage{multirow}
\usepackage{tabularx}

\usepackage[sort&compress]{natbib}

\begin{document}

\title{Electromagnetically induced transparency and wide-band wavelength conversion in silicon nitride microdisk optomechanical resonators}

\author{Yuxiang Liu}\homepage{These authors contributed equally}
\affiliation{Center for Nanoscale Science and Technology,
National Institute of Standards and Technology, Gaithersburg, MD
20899, USA}
\affiliation{Institute for Research in Electronics and Applied Physics, University of Maryland, College Park, MD 20742, USA}
\author{Marcelo Davan\c co}\homepage{These authors contributed equally}
\affiliation{Center for Nanoscale Science and Technology,
National Institute of Standards and Technology, Gaithersburg, MD
20899, USA}
\affiliation{Department of Applied Physics, California Institute of Technology, Pasadena, CA 91125, USA}
\author{Vladimir Aksyuk}
\affiliation{Center for Nanoscale Science and Technology, National
Institute of Standards and Technology, Gaithersburg, MD 20899, USA}
\author{Kartik Srinivasan}\email{kartik.srinivasan@nist.gov}
\affiliation{Center for Nanoscale Science and Technology, National
Institute of Standards and Technology, Gaithersburg, MD 20899, USA}
\date{\today}

\begin{abstract}
We demonstrate optomechanically-mediated electromagnetically-induced transparency and wavelength conversion in silicon nitride (Si$_3$N$_4$) microdisk resonators.  Fabricated devices support whispering gallery optical modes with a quality factor ($Q$) of 10$^6$, and radial breathing mechanical modes with a $Q$=10$^4$ and a resonance frequency of 625~MHz, so that the system is in the resolved sideband regime.  Placing a strong optical control field on the red (blue) detuned sideband of the optical mode produces coherent interference with a resonant probe beam, inducing a transparency (absorption) window for the probe.  This is observed for multiple optical modes of the device, all of which couple to the same mechanical mode, and which can be widely separated in wavelength due to the large bandgap of Si$_3$N$_4$.  These properties are exploited to demonstrate frequency upconversion and downconversion of optical signals between the 1300 nm and 980 nm bands with a frequency span of 69.4~THz.

\end{abstract}


\maketitle
Recent demonstrations of strong radiation pressure interactions in cavity optomechanics have focused on the resolved sideband regime, where mechanical sidebands of the optical mode lie outside its linewidth~\cite{ref:Kippenerbg_Vahala_Science}.  Such systems have been used in laser cooling a mechanical oscillator to its ground state~\cite{ref:Teufel_ground_state,ref:Chan_Painter_ground_state}, coherent interference effects such as electromagnetically-induced transparency (EIT)~\cite{ref:Weis_Kippenberg,ref:safavi-naeini4}, parametrically-driven normal mode splitting~\cite{ref:Groblacher_Aspelmeyer_Nature,ref:Teufel_strong_coupling,ref:Verhagen_Kippenberg_Nature}, and observing energy exchange between the optical and mechanical systems~\cite{ref:Fiore_Wang_PRL,ref:Verhagen_Kippenberg_Nature}.  Here, we study a system consisting of a small diameter silicon nitride (Si$_3$N$_4$) microdisk in which multiple high quality factor optical modes couple to a 625~MHz mechanical radial breathing mode.  We demonstrate optomechanically-mediated EIT and wavelength conversion~\cite{ref:Tian_Wang_PRA,ref:safavi-naeini3,ref:Wang_Clerk_state_transfer_PRL,ref:Hill_Painter_WLC_Nat_Comm,ref:Dong_Wang_Science_dark_mode}, up- and downconverting signals across the widely separated 1300~nm and 980~nm wavelengths bands.  Our results establish Si$_3$N$_4$ as a viable platform for chip-scale cavity optomechanics in the resolved sideband regime.  More generally, Si$_3$N$_4$ offers potential integration of cavity optomechanics with numerous classical and quantum photonic elements, including ultra-low-loss passive components~\cite{ref:Bauters_Bowers_Si3N_wgs}, microcavity  frequency combs~\cite{ref:Levy_Lipson_comb} and spectrally narrow mode-locked lasers~\cite{ref:Peccianti_Morandotti_NComm}, and integrated superconducting single photon detectors~\cite{ref:Schuck_Tang_Si3N4_SSPD}.

In the context of chip-scale guided wave devices, experiments making use of frequency-resolved mechanical sidebands have largely been in two systems exhibiting significantly different parameter regimes: silica microtoroid cavities~\cite{ref:Weis_Kippenberg,ref:Verhagen_Kippenberg_Nature} and silicon photonic and phononic crystal resonators (optomechanical crystals)~\cite{ref:eichenfield2,ref:safavi-naeini4,ref:Chan_Painter_ground_state}. Silica microtoroids support ultra-high quality factor optical modes ($Q_{\text{o}}$$>$$10^7$, decay rate $\kappa/2\pi$$\approx$10~MHz) that are coupled to $\gtrsim$50~MHz frequency mechanical modes with a zero-point optomechanical coupling rate $g_{\text{0}}/2\pi$$\gtrsim$1~kHz.  In contrast, silicon optomechanical crystals have $Q_{\text{o}}$$\approx10^6$ ($\kappa/2\pi$$\gtrsim$200~MHz) modes that couple to few GHz mechanical modes with $g_{\text{0}}/2\pi$$\approx$1~MHz.  Though similar physics has been studied in both, there are qualitative benefits in each system.  Higher mechanical frequencies yield lower phonon occupation numbers for a given temperature, and larger bandwidths (for a given mechanical quality factor $Q_{\text{m}}$) in phenomena like slow light~\cite{ref:safavi-naeini4} and wavelength conversion~\cite{ref:Hill_Painter_WLC_Nat_Comm}.  Wide bandgap materials like silica enable operation across a broad wavelength range, including visible wavelengths common to atomic systems, and are free of nonlinear loss mechanisms at most wavelengths.  Here, we develop Si$_3$N$_4$ microdisks as a chip-scale cavity optomechanical platform that combines many advantageous features of the aforementioned systems.  This includes near-GHz mechanical frequencies, a high frequency-$Q_{\text{m}}$ product (6$\times$10$^{12}$~Hz), a straightforward optical mode structure with multiple high-$Q_{\text{o}}$ ($\approx$$10^6$) optical modes coupled to the same mechanical mode ($g_{\text{0}}/2\pi$$\approx$8~kHz), a wide bandgap with broad optical transparency, and low nonlinear loss. Although microdisk optomechanical devices have been used in sensitive optical transduction of motion~\cite{ref:Anetsberger_near_field,*ref:Srinivasan_Nano_Letters} (including at GHz frequencies~\cite{ref:Ding_Favero_GaAs_disk_optomechanics,*ref:Sun_Tang_GHz_microdisk}), mechanical oscillation and cooling~\cite{ref:Lin5}, and optical mode tuning~\cite{ref:Wiederhecker_Lipson}, we note the qualitative difference with respect to our experiments, which demonstrate coherent interference phenomena like EIT and wavelength conversion.

\begin{figure}
\centerline{\includegraphics[width=8.5 cm, height=6.3 cm]{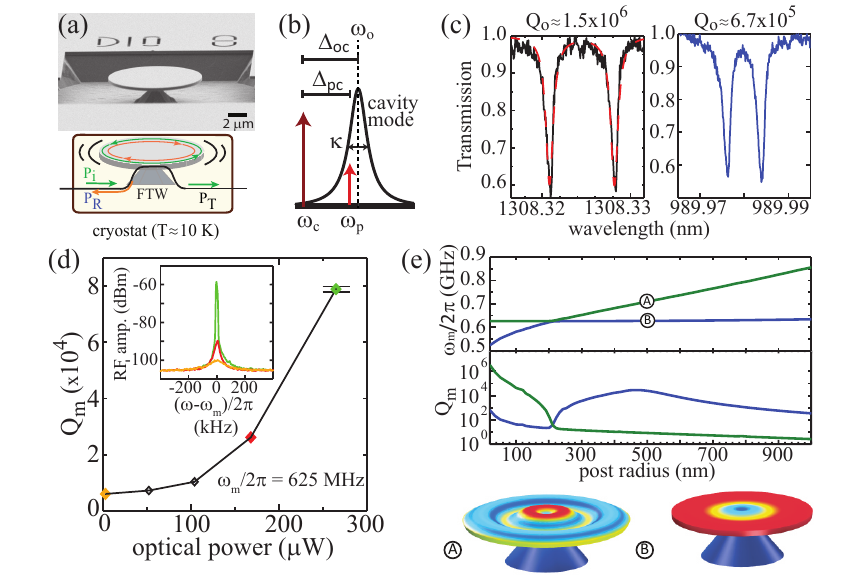}}
\caption{(a) Scanning electron micrograph and schematic of the fiber taper waveguide (FTW) coupling to the disk, with optical (green and orange curved arrows) and mechanical (black curved lines) modes depicted. (b) Sideband spectroscopy: $\omega_{\text{o,c,p}}$ are the frequencies of the optical cavity mode, control field, and probe field, respectively, and $\kappa$ is the cavity mode linewidth.  (c) Room-temperature 1300 nm and 980~nm band optical modes. (d) Room-temperature mechanical $Q$ as a function of input optical power (laser blue-detuned from the optical mode) for the radial breathing mode.  The inset shows the mechanical mode spectrum at three highlighted optical powers. (e) Finite-element-method calculated frequency (top) and mechanical $Q$ (middle) for two modes, labeled \textcircled{\hspace{0.008in}\footnotesize{A}} and \textcircled{\hspace{0.008in}\footnotesize{B}}, as a function of top pedestal radius. The modes anti-cross at a radius of 200~nm.  (bottom) Displacement profiles for a radius of 500~nm.} \label{fig:fig1}
\vspace{-0.2in}
\end{figure}

Microdisk cavities (Fig.~\ref{fig:fig1}(a)) are fabricated in 350~nm thick Si$_3$N$_4$-on-silicon, as described in the Supplementary Material~\cite{ref:Microdisk_EIT_note}. The choice of geometry is informed by finite element simulations~\cite{ref:Microdisk_EIT_note}.  The disk diameter (10~$\mu$m) is small enough to support a high frequency ($>0.6$~GHz) mechanical radial breathing mode, and large enough to avoid intrinsic optical radiation loss.  The top pedestal diameter (200~nm) limits coupling between the radial breathing mode and lossy mechanical modes of the supporting pedestal (Fig.~\ref{fig:fig1}(e)). Basic optical and mechanical properties are experimentally characterized using swept-wavelength spectroscopy, with light coupled into the devices using an optical fiber taper waveguide (FTW).  At room temperature and under moderate vacuum ($0.002$~Pa $\approx$~2$\times$10$^{-5}$~torr), an optical quality factor $Q_{\text{o}}$ as high as 2$\times$10$^6$ at both 1300~nm and 980~nm is measured. Figure~\ref{fig:fig1}(c) shows representative scans for a device whose total $Q_{\text{o}}$ is 1.5$\times$10$^6$ at 1308~nm, and 6.7$\times$10$^5$ at 990~nm (the intrinsic $Q_{\text{o}}$, based on the depth of coupling, is 1.7$\times$10$^6$ and 7.6$\times$10$^5$, respectively). Mechanical modes are measured by tuning the laser to the shoulder of an optical mode and detecting optical field fluctuations induced by the thermally-driven motion of the disk.  We observe the fundamental radial breathing mode at $\omega_{m}/2\pi\approx625$~MHz, with undriven $Q_{\text{m}}$ of $6\times$10$^3$ (Fig.~\ref{fig:fig1}(d)).  To improve thermal stability, the device is cooled in a liquid He cryostat to $\approx10$~K. We see an increase in the undriven $Q_{\text{m}}$ to 1$\times$10$^4$, yielding a frequency-$Q_{\text{m}}$ product of 6$\times$10$^{12}$~Hz, which approaches the recently demonstrated value of 2$\times$10$^{13}$~Hz in Si$_3$N$_4$ (albeit at a much lower frequency of 10~MHz~\cite{ref:Wilson_Kimble_PRL}).  Cryogenic operation also tends to degrade $Q_{\text{o}}$ (by as much as a factor of 3), which we attribute to cryo-gettering of material on the sample.  The splitting in the doublet mode optical transmission spectra, which results from backscattering~\cite{ref:Weiss}, also changes.  These changes vary from cooldown to cooldown.

In the context of cavity optomechanics, EIT is an increase in the transmission of a near-resonant probe beam (at frequency $\omega_{\text{p}}$) through an optical cavity mode (at frequency $\omega_{\text{o}}$) that results when a strong control field (at frequency $\omega_{\text{c}}$) is red-detuned by $\Delta_{\text{oc}}=\omega_{\text{o}}-\omega_{\text{c}}=\omega_{\text{m}}$ (Fig.~\ref{fig:fig1}(b)), so that anti-Stokes photons generated by scattering of the control field by the mechanical resonator interfere with the probe and create a transparency window for it~\cite{ref:Weis_Kippenberg,ref:safavi-naeini4,ref:Agarwal_EIT}.  We focus on the reflected signal from the cavity, present in our microdisks due to the aforementioned backscattering, and for which EIT results in a narrow dip in the reflection peak.  Our measurement setup is described in detail in the Supplementary Material~\cite{ref:Microdisk_EIT_note}.

\begin{figure*}
\centerline{\includegraphics{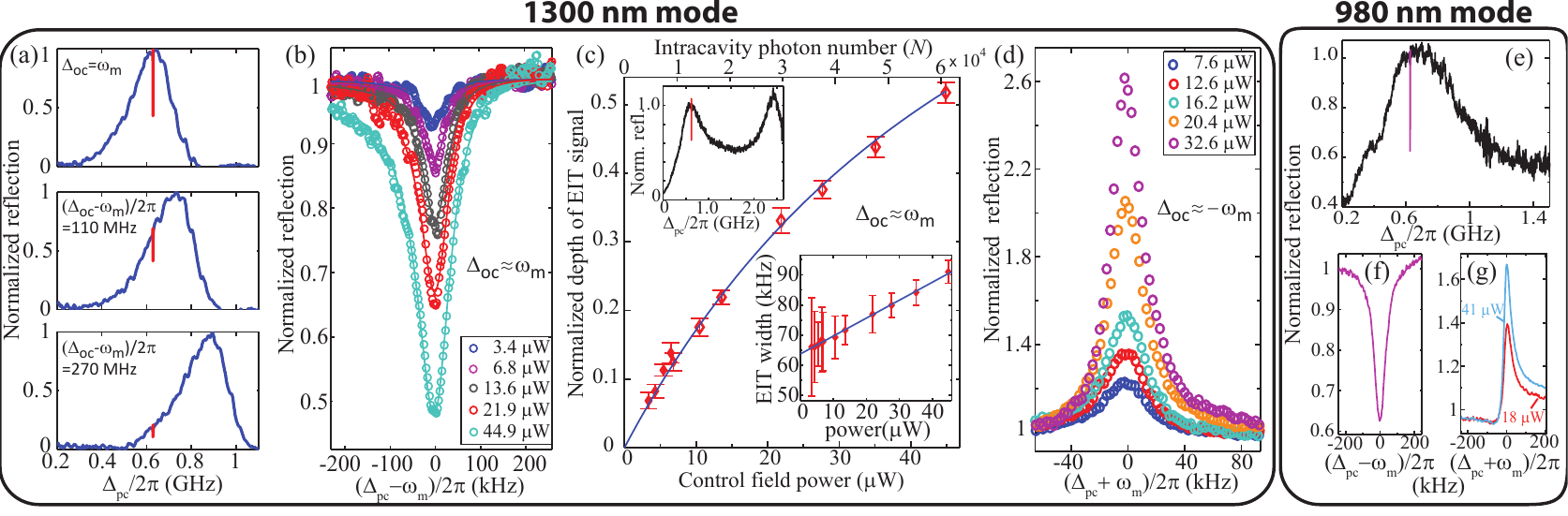}}
\caption{EIT/EIA measurements for a (a)-(d) 1300~nm and (e)-(g) 980~nm mode. (a) Reflection spectra with the optical mode-control field detuning ($\Delta_{\text{oc}}$) stepped and the probe-control detuning ($\Delta_{\text{pc}}$) swept.  Blue (red) curves are taken over a broad (narrow) range of $\Delta_{\text{pc}}$. (b) Zoomed-in reflection spectra around the EIT dip (open circles are data, solid lines are fits) and (c) Depth of the EIT dip as a function of control field power, for $\Delta_{\text{oc}}\approx\omega_{m}$.  The upper inset shows the broad reflection spectrum corresponding to the data in (b)-(c), for which $Q_{\text{o}}$ had degraded with respect to that in (a). The lower inset shows the width of the EIT dip in (b) as a function of control field power. (d) Zoomed-in reflection spectra for $\Delta_{\text{oc}}=-\omega_{m}$, for which EIA is observed. (e) Broad, and (f)-(g) narrow reflection spectra as a function of $\Delta_{\text{pc}}$ for a cavity mode in the 980~nm band.  In (e)-(f), $\Delta_{\text{oc}}\approx\omega_{m}$, while in (g), $\Delta_{\text{oc}}\approx-\omega_{m}$.}\label{fig:fig2}
\vspace{-0.2in}
\end{figure*}

We first study the system in the 1300~nm band, by sweeping $\Delta_{\text{pc}}=\omega_{\text{p}}-\omega_{\text{c}}$ at different values of $\Delta_{\text{oc}}$.  For each $\Delta_{\text{oc}}$, we take two sweeps of $\Delta_{\text{pc}}$, one over a broad range to trace the overall cavity reflection spectrum and lock the control laser, and the other over a narrow range to resolve the narrow EIT dip~\cite{ref:Microdisk_EIT_note}.  A series of spectra is shown in Fig.~\ref{fig:fig2}(a), where $Q_{\text{o}}$ has slightly degraded compared to its room temperature value, with $\kappa/2\pi$$\approx$180~MHz.  As observed in previous work~\cite{ref:Weis_Kippenberg,ref:safavi-naeini4}, the EIT dip always appears at $\Delta_{\text{pc}}$=$\omega_{\text{m}}$, and its width is given by the total mechanical damping rate $\gamma$, which is the sum of the intrinsic rate $\gamma_{m}$ and the optomechanically-induced damping rate $\gamma_{\text{OM}}=C\gamma_{m}$, where $C$ is the cooperativity parameter. $C=\frac{4G^2}{\kappa\gamma_{m}}$, where $G=g_{\text{0}}\sqrt{N}$ is the parametrically-enhanced coupling rate provided by $N$ intracavity control field photons.  We reach $C$$\approx$0.45 by increasing the optical power before the system is thermally unstable.

To more quantitatively understand the EIT effect and assess the zero-point optomechanical coupling rate $g_{\text{0}}$, we set $\Delta_{\text{oc}}$$\approx$$\omega_{\text{m}}$ and measure the normalized reflection spectrum as a function of control field power (Fig.~\ref{fig:fig2}(b)).  For this set of measurements, $Q_{\text{o}}$ degraded upon cooldown to $\kappa/2\pi$$\approx$550~MHz (inset to Fig.~\ref{fig:fig2}(c)), while the mechanical mode had $\gamma_{m}/2\pi$$\approx$63~kHz.  We fit each normalized reflection spectrum using the theory presented in Ref.~\onlinecite{ref:safavi-naeini4}, where the amplitude reflection coefficient $r(\Delta_{\text{pc}})$, normalized to unity, is given as:

\vspace{-0.2in}
\begin{equation}
r(\Delta_{\text{pc}}) = -\frac{1}{1+\frac{2i(\Delta_{\text{oc}}-\Delta_{\text{pc}})}{\kappa}+\frac{C}{\frac{2i(\omega_{\text{m}}-\Delta_{\text{pc}})}{\gamma_{\text{m}}}+1}}
\label{eq:eq1}
\end{equation}
\vspace{-0.15in}

\noindent $R(\Delta_{\text{pc}})$=$|r|^2$ is the normalized reflected intensity, and the contrast in the EIT dip, $\Delta R=1-R(\Delta_{\text{pc}}=\omega_{\text{m}})$, is plotted as a function of control field power in Fig.~\ref{fig:fig2}(c) (in the limit $\Delta_{\text{oc}}$=$\omega_{\text{m}}$, $\Delta R =1-1/(1+C)^2$). Recalling that $C=\frac{4Ng_{\text{0}}^2}{\kappa\gamma_{m}}$, we use Eq.~(\ref{eq:eq1}) in a fit to extract $g_{\text{0}}/2\pi$$\approx$7.8~kHz, where knowledge of the FTW loss, cavity mode spectrum, and $\Delta_{\text{oc}}$ allows us to estimate $N$ at each pump power~\cite{ref:Microdisk_EIT_note}.  We see good agreement between experiment and theory, further supported by the linear increase in EIT dip width with pump power (lower inset to Fig.~\ref{fig:fig2}(c)).  However, the experimentally determined $g_{\text{0}}$ is nearly two times smaller than that predicted from finite element simulations ($g_{\text{0}}/2\pi$$\approx$15~kHz) that exclusively consider the contribution due to moving dielectric boundaries.  A study of whether photo-elastic effects may play a role~\cite{ref:Chan_Painter_optimized_OMC} in this discrepancy is underway.  Finally, shifting the control field to the blue-detuned side of the optical mode ($\Delta_{\text{oc}}$$\approx$$-\omega_{\text{m}}$) results in electromagnetically induced absorption (EIA), evidenced by a peak in the center of the reflection spectrum that increases with optical power (Fig.~\ref{fig:fig2}(d)).

Whispering gallery mode cavities can support high-$Q$ optical modes over a broad range of wavelengths.  As a result, multiple, broadly spaced optical modes may be expected to couple to the same mechanical mode, which we observe in our measurements.  Figure~\ref{fig:fig2}(e)-(g) presents EIT and EIA data for a 980~nm band optical mode ($\kappa/2\pi$$\approx$750~MHz) that couples to the $625$~MHz radial breathing mode.  We observe similar behavior as seen for the 1300~nm band mode, though the maximum contrast of the EIT dip and EIA peak are a little smaller, likely due to lower $Q_{\text{o}}$ which is not fully compensated by higher optical power before thermal instability sets in.

\begin{figure*}
\begin{center}
\includegraphics[width=\linewidth]{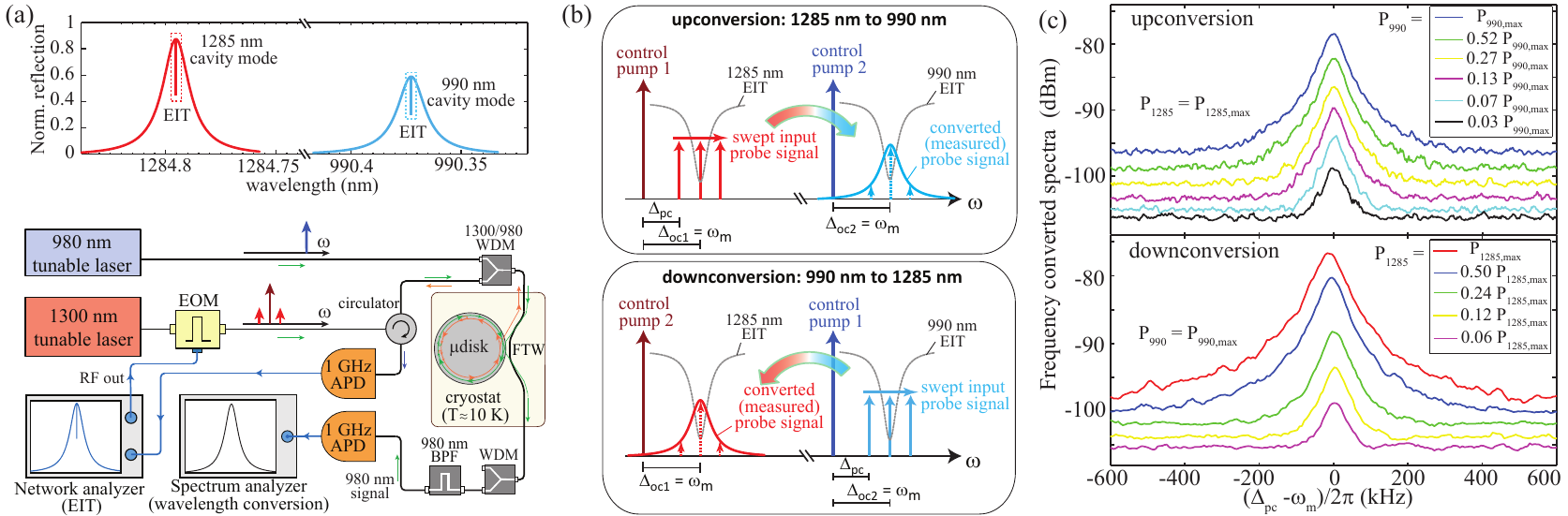}
\caption{Wide-band wavelength conversion: (a)-(b) Schematics of the process. (a) Top: Frequency conversion occurs over a pair of transparency windows opened up in two of the cavity's optical modes through EIT. Bottom: Experimental setup for upconversion. The frequency converted signal transmitted past the cavity is measured on a RF spectrum analyzer. (b) Zoomed-in schematics of the EIT transparency windows in (a). Top: In frequency upconversion, a 1285~nm control pump and 990~nm control pump are used to convert a 1285~nm input probe signal to 990~nm.  The detuning between the 1285~nm input probe signal and control pump can be swept over a range given by the bandwidth of the damped mechanical oscilator. (Bottom) In frequency downconversion, the input probe signal is at 990~nm, and the control pumps mediate conversion to 1285~nm. (c) Upconverted (top) and downconverted (bottom) signals, measured as the relevant input probe signal-control pump detuning ($\Delta_{\text{pc}}$) is swept.  In upconversion (downconversion), the 1285~nm (990~nm) control pump power is fixed at $P_{\text{1285,max}}$=63.8~$\mu$W ($P_{\text{990,max}}$=9.0~$\mu$W) and the 990~nm (1285~nm) control pump power is stepped.}
\label{fig:fig3}
\end{center}
\vspace{-0.2in}
\end{figure*}

One application of multiple optical modes coupled to the same mechanical mode is in wavelength conversion, as outlined in theory~\cite{ref:safavi-naeini3,*ref:Wang_Clerk_state_transfer_PRL} and recently demonstrated in silicon optomechanical crystals~\cite{ref:Hill_Painter_WLC_Nat_Comm} and silica microspheres~\cite{ref:Dong_Wang_Science_dark_mode}. The application of two control pumps, each red-detuned from a corresponding optical mode by $\omega_{\text{m}}$, opens up a pair of transparency windows over which wavelength conversion mediated by the mechanical resonator can occur (Fig.~\ref{fig:fig3}(a)-(b)).  Input signals over a bandwidth set by the damped mechanical resonator can be upconverted or downconverted, with an internal conversion efficiency that depends on the cooperativity achieved for each optical mode.

We proceed following the recent experiments of Hill et~al.~\cite{ref:Hill_Painter_WLC_Nat_Comm}, with a simplified experimental setup shown in Fig.~\ref{fig:fig3}(a) and described further in the Supplementary Material~\cite{ref:Microdisk_EIT_note}. We first alternate between EIT spectroscopy in the 1300~nm and 980~nm bands.  These measurements are done to ascertain $\Delta_{\text{oc}}$ for each mode, as well as the cooperativity achieved. For the experiments that follow, $\Delta_{\text{oc}}$$\approx$$\omega_{\text{m}}$ for both control pumps. Focusing first on frequency upconversion, we amplitude modulate the 1300~nm laser to generate an input probe signal that is detuned from the control pump by $\Delta_{\text{pc}}$. The control pump and input probe signal in the 1300~nm band are combined with the control pump in the 980~nm band and sent into the FTW-coupled microdisk.  Light exiting the device is then spectrally separated into the 980~nm and 1300~nm bands. As the converted tone in the 980~nm band is detuned with respect to the 980~nm band control pump by $\Delta_{\text{pc}}$, the beating of the two fields is recorded on an electronic spectrum analyzer after photodetection.  The input probe-control field detuning $\Delta_{\text{pc}}$ is then swept to assess the conversion bandwidth (Fig.~\ref{fig:fig3}(b)).

The Supplementary Material~\cite{ref:Microdisk_EIT_note} shows the optical transmission spectrum measured for modes in the 1300~nm and 980~nm bands. The top of Fig.~\ref{fig:fig3}(c) shows a series of measurements for frequency upconversion, in which the power in the 1285~nm control pump is held fixed at its maximum value, and the power in the 990~nm control pump is stepped. Each displayed trace, taken at fixed control pump powers, represents the envelope of a series of frequency converted spectra measured on the RF spectrum analyzer, in which the frequency separation between the input probe signal and adjacent control pump $\Delta_{\text{pc}}$ is swept.  This provides us with an estimate of the bandwidth of the conversion process, which matches that of the damped mechanical oscillator.  As the power is increased towards its highest values, the conversion efficiency (the photon number ratio of the converted signal to input signal) begins to saturate, with internal (external) conversion efficiencies reaching $\approx16.1~\%$ ($0.49~\%$)~\cite{ref:Microdisk_EIT_note}.

The maximum expected internal conversion efficiency is estimated from the cooperativities of the two modes, $C_{1}$ and $C_{2}$~\cite{ref:Hill_Painter_WLC_Nat_Comm,ref:Dong_Wang_Science_dark_mode}, as $\eta=4C_{1}C_{2}/(1+C_{1}+C_{2})^2$. While in Fig.~\ref{fig:fig2}, $C_{1}$=$C_{2}$=$0.45$ was reached, corresponding to a maximum $\eta=22~\%$, the lower $Q_{\text{o}}$s measured here caused a reduction in $C_{1}$ and $C_{2}$ that was not fully compensated by additional optical power, thus explaining the lower efficiency. The external conversion efficiency of $0.49\%$ is given by the product of the internal efficiency and the waveguide-cavity incoupling and outcoupling efficiencies, which are 29.2~$\%$ and 10.3~$\%$, respectively.

For frequency downconversion, the roles of the 990~nm and 1285~nm lasers are reversed (Fig.~\ref{fig:fig3}(b)), as now an input probe signal at 990~nm is combined with control pumps in the 980~nm and 1300~nm bands, to generate converted tones in the 1300~nm band.  The resulting downconversion data measured on the RF spectrum analyzer is shown in the bottom of Fig.~\ref{fig:fig3}(c).  This data is taken by fixing the power in the 990~nm control pump at its maximum value, and stepping the power in the 1285~nm control pump.  Each trace is the envelope of a series of frequency converted spectra, where the detuning between input probe signal and adjacent control pump $\Delta_{\text{pc}}$ is swept.

Eventually, such wavelength conversion could connect quantum optical technology at wavelengths below 1000~nm (e.g., quantum memories) and at 1300~nm (low loss/dispersion optical fibers). Several improvements must be made for such quantum frequency conversion~\cite{ref:Kumar} to be feasible.  Higher cooperativities and more efficient waveguide-cavity coupling are needed to improve the conversion efficiency.  Reduced mechanical damping is one key to increased cooperativity. As simulations (Fig.~\ref{fig:fig1}(e)) indicate that coupling to pedestal modes may limit $Q_{\text{m}}$, smaller pedestal diameters or isolation of the disk periphery from a central, pedestal-supported region~\cite{ref:Anetsberger_Kippenberg_low_dissipation} may provide improvement.

The generation of noise photons is a crucial consideration for applications such as frequency conversion of single photon states~\cite{ref:Raymer_Srinivasan_PT}, and has contributions that stem from the limited sideband resolution of the system (e.g., Stokes scattered photons) and the thermal occupancy of the mechanical resonator ($\langle n \rangle_{\text{th}}$$\approx$330 for the mechanical mode at 10~K; laser cooling in our experiments reduces this by at most $\approx$$33$$~\%$).  In contrast, recent work in optomechanical crystals~\cite{ref:Hill_Painter_WLC_Nat_Comm} cooled the system close to the ground state ($\langle n \rangle_{\text{th}}$$\approx$3), whereas experiments done at room temperature in microspheres~\cite{ref:Dong_Wang_Science_dark_mode} have an even larger noise contribution ($\langle n \rangle_{\text{th}}$$\approx$4$\times$10$^4$).

Finally, Si$_3$N$_4$ devices may benefit by supporting a large intracavity photon number $N$ without nonlinear loss which, for example, influences Si optomechanical crystals at $N$$\gtrsim$300~\cite{ref:Chan_Painter_ground_state}. We reach $N$$\approx$6$\times$10$^4$, and further increases are not limited by nonlinear loss, but instead thermal stability.  Techniques to lock $\Delta_{\text{oc}}$ beyond the relatively slow procedure we have adopted~\cite{ref:Microdisk_EIT_note} would be of significant benefit.  Finally, increased $g_{\text{0}}$ could dramatically improve performance, given the squared dependence of the cooperativity on this parameter.  While large increases in $g_{\text{0}}$ are unlikely for microdisks, since significant decreases in diameter will result in radiation losses, recent designs of Si$_3$N$_4$ slot mode optomechanical crystals~\cite{ref:Davanco_slot_mode_OMC} suggest that such a system may be able to combine high frequency, large $g_{\text{0}}$, and large $N$.

We thank O. Painter for helpful discussions and loan of equipment.  Y.L. acknowledges support under the NIST-ARRA Measurement Science and Engineering Fellowship Program Award 70NANB10H026 through the University of Maryland. This work was partly supported by the DARPA/MTO MESO program.

\vspace{-0.3in}

\newpage

\section{Supplementary Material}

\section{Finite element simulations}

We performed finite element method simulations of our microdisk structures to gain a clearer physical understanding of the optomechanical system.

\subsection*{Obtaining optical and mechanical cavity modes}

Our microdisk cavities support whispering-gallery optical modes (WGMs) of the form $\mathbf{E}(r,z)\exp\left(im\phi\right)$, where $\mathbf{E}(r)$ is the electric field distribution on the $rz$ plane (in cylindrical spatial coordinates) and $m$ is the azimuthal order. Such WGMs were calculated by solving the 2D axially-symmetric electromagnetic wave equation
\begin{equation}
\nabla\times\nabla\times\mathbf{E} =\epsilon(\mathbf{r})\left(\frac{\omega}{c}\right)^2\mathbf{E}.
\label{eq:fem1}
\end{equation}
via the finite element method, with a formulation that employed respectively edge and node elements for the transverse and longitudinal electric field components, and perfectly-matched layers to simulate open boundaries. Solving the eigenvalue problem in eq.~(\ref{eq:fem1}) for a specific azimuthal order $m$ produced eigenmodes with complex frequencies $\omega_m$ and optical quality factors $Q_m ={\text{Re}}\{\omega_m\}/(2{\text{Im}}\{\omega_m\})$. Because WGM fields are mostly concentrated in the periphery of the resonator, interaction with the Si pedestal that supports the SiN microdisk is negligible, and thus completely ignored in the optical mode calculation.

For disk radius $D\approx10~\mu$m and SiN thickness $t\approx350$~nm, modes with radiation-limited quality factors exceeding $10^8$ can be found both in the 980 nm and 1300 nm bands. Table~\ref{tab:1} shows wavelengths and quality factors of the calculated TE and TM polarized WGMs, which correspond well with experiment. The good agreement between the experimental and calculated values was achieved by tuning the parameters of the cavity within reasonable bounds (thickness $t=340$~nm and $n = 1.99$ for the refractive index of SiN), compared to experimentally estimated values.

\begin{table}[h!]
\setlength{\extrarowheight}{2pt}
\caption{ Calculated and experimental whispering gallery modes} \centerline{
    \begin{tabular}{|c|c|c|c|c|c|c|}
        \hline
        $\lambda_{\text{FEM}}$ (nm) & $Q_{\text{o,FEM}}\times10^8$ & Polarization  & $\lambda_{\text{exp.}}$ (nm)  \\ \hline
        \hline
        971.28 & 10 & TE & 970  \\
        987.37 & 4.5 & TE & 984  \\
        975.36 & 5.7 & TM & 974  \\
        989.61 & 2.4 & TE & 990  \\
        1281.23 & 1.0 & TE & 1285  \\
        1309.04 & 0.5 & TE & 1308  \\
        \hline
    \end{tabular}
}    \label{tab:1}
\end{table}

The mechanical modes of the microdisk structure were obtained by solving the equation of motion for the displacement field $\mathbf{Q}(\mathbf{r})$, assuming anisotropic materials~\cite{ref:eichenfield2}.
In our simulations,  the Si pedestal supporting the SiN microdisk was represented by a conical frustum whose angle and height were equivalent to those determined from scanning electron microscope images of fabricated structures. A zero displacement boundary condition was enforced at the bottom surface of the conical frustum, corresponding to the region where the pedestal meets the Si substrate.

For the $10~\mu$m diameter disks considered, a first-order radial breathing mode (RBM) is obtained in the vicinity of $f_m$=625~MHz. The displacement profile of the RBM, as shown at the bottom of Fig. 1(e) (mode \textcircled{\footnotesize{B}}), is primarily in the radial direction and confined to the disk plane, with relatively small vertical displacement in the disk-pedestal contact area. Due to its azimuthal symmetry, the RBM is expected to display preferential optomechanical coupling to the microdisk's WGMs.

\subsection*{Optomechanical coupling}
The shift in the frequency $\omega_{o}$ of a
particular optical resonance due to displacement of the
nanostructure boundaries produced by a mechanical resonance at
frequency $f_m$ is quantified by the optomechanical coupling $g_{om} =
\partial \omega_o/\partial x=\omega_o/L_{\text{OM}}$; here, $x$ is the cavity
boundary displacement and $L_{\text{OM}}$ is an effective optomechanical interaction
length~\cite{ref:eichenfield2}. The effective length $L_{\text{OM}}$ can be
estimated via the perturbative expression
\begin{equation}
\label{eq:Lom} L_{\text{OM}} = \frac{2\int {dV \epsilon\left|\mathbf{E}
\right|^2}}{\int{dA\left(\mathbf{Q}\cdot \mathbf{n}
\right)\left( \Delta \epsilon\left| \mathbf{E}_\parallel
\right|^2-\Delta(\epsilon^{-1}) \left| \mathbf{D}_\perp \right|
^2\right)} }.
\end{equation}

Here, $\mathbf{E}$ and $\mathbf{D}$ are the modal electric and
electric displacement fields, respectively, $\Delta \epsilon =
\epsilon_{diel.}-\epsilon_{air}$, $\Delta(\epsilon^{-1}) =
\epsilon_{diel.}^{-1}-\epsilon_{air}^{-1}$, and $\epsilon_{diel.}$ and $\epsilon_{air}$ are the permittivities of the microdisk material and air, respectively. The
mass displacement due to the mechanical resonance is given by
$\mathbf{Q}$, and the normal surface displacement at the structure
boundaries is $\mathbf{Q}\cdot \mathbf{n}$, where
$\mathbf{n}$ is the surface normal. The integral in the denominator
is performed over the entire surface of the nanostructure.

The optomechanical coupling $g_{om}$ can be converted into a pure
coupling rate $g_0$ between the optical and mechanical resonances, with
$g_0=x_{zpf}\cdot g_{om}$, where $x_{zpf}=\sqrt{\hbar/2m\omega_m}$ is
the zero point fluctuation amplitude for mechanical
displacement and $m$ is the motional mass
of the mechanical resonance at frequency $\omega_m$. The motional
mass can be obtained from the displacement $\mathbf{Q}$ and the
nanobeam material density $\rho$ by
$m=\rho\int{dV\left(\frac{|\mathbf{Q}|}{\text{max}(|\mathbf{Q}|)}\right)^2}$~\cite{ref:safavi-naeini3}.

For the RBM at $f_m\approx625$~MHz above, $m\approx60$~pg and $x_{zpf}\approx0.5$~fm. Table~\ref{tab:2} shows calculated values of $L_{\text{om}}$, $g_{om}$ and the zero-point coupling rate $g_0$ between the RBM and TE-polarized WGMs in both the 980~nm and 1300~nm bands. The calculated $g_{om}$ values are smaller than those estimated via the expression $g_{om}=\omega_{o}/R$, with $\omega_{o}$ the optical mode frequency and R the disk radius, which is commonly employed for the case of large disks. For smaller radius disks such as considered here, optical fields extend considerably into the vacuum regions surrounding the dielectric, resulting in an effective optomechanical length $L_{\text{om}}$ larger than the actual physical dimension $R$, as seen in Table~\ref{tab:2}.

\begin{table}
\setlength{\extrarowheight}{2pt}
\caption{ Optomechanical coupling parameters between 980 nm and 1300 nm TE-polarized WGMs and 625 MHz RBM} \centerline{
    \begin{tabular}{|c|c|c|c|c|c|c|}
        \hline
        $\lambda_{\text{o}}$ (nm) & $Q_{\text{o}}\times10^8$ & $g_{om}/2\pi$ (GHz/nm) & $g_0/2\pi$ (kHz) & $L_{\text{om}}$ ($\mu m$) \\ \hline
        \hline
        988.41 & 4.8 & 42.7 & 19.9 & 7.10 \\ \hline
        1281.80 & 1.0 & 33.2 & 15.5 & 7.05 \\ \hline
    \end{tabular}
}    \label{tab:2}
\end{table}

\subsection*{Clamping losses}
To estimate mechanical clamping losses of the mechanical resonances, we adopted the method of ref.~\cite{ref:Anetsberger_Kippenberg_low_dissipation}, in which the contact region between the disk and the pedestal is modeled as a membrane that radiates acoustic energy with power
\begin{equation}
P = c\rho\Omega_m^2\int_AdA\left|\mathbf{Q(r)}\cdot\mathbf{\hat z}\right|^2,
\end{equation}
where $\rho$ is the density of the disk material and $c$ is the speed of sound, $\Omega_m$ is the angular frequency of the mechanical resonance, and $\left|\mathbf{Q(r)}\cdot\mathbf{\hat z}\right|^2$ is the out-of-disk-plane displacement over the contact area $A$. If $W_{mech}$ is the stored mechanical energy of the resonance, the latter's mechanical quality factor can be estimated as
\begin{equation}
Q_{\text{m}} = \left(\frac{P}{\Omega_mW_{mech}}\right)^{-1}.
\end{equation}

\begin{figure}[b!]
\centerline{\includegraphics[width=6 cm]{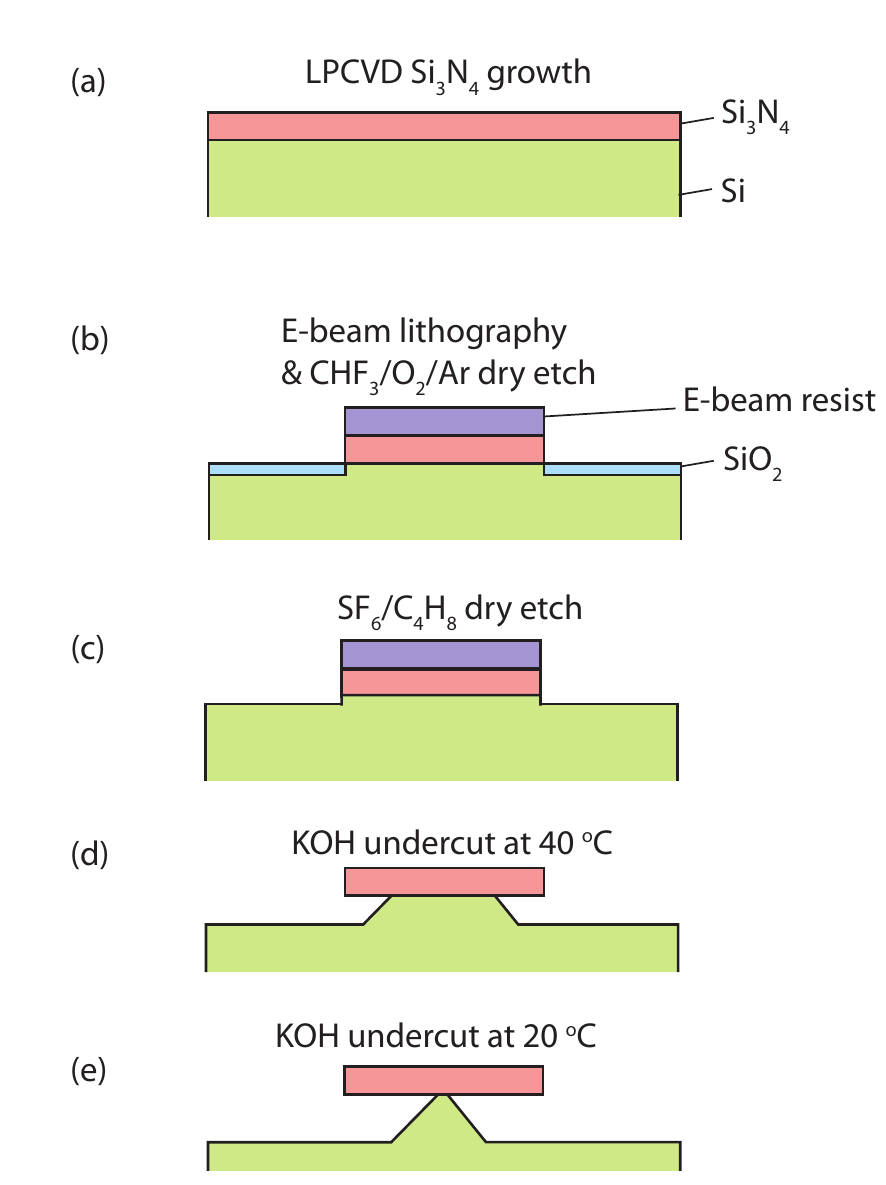}}
\caption{Processing steps for fabrication of the Si$_3$N$_4$ microdisks.}
\label{fig:SFig1}
\end{figure}

Assuming a conical frustum for the pedestal shape, we calculate mechanical modes as a function of pedestal radius at the interface with Si$_3$N$_4$ (Fig.~1(e) top), with displacement amplitude profiles shown for two modes at a radius of 500~nm  (Fig.~1(e) bottom). We also plot the mechanical mode frequencies and $Q$s due to clamping losses, following the approach of Anetsberger et~al.~\cite{ref:Anetsberger_Kippenberg_low_dissipation}. It is apparent in Fig.~1(e) that, for large top pedestal radius, the mechanical quality factor for the RBM tends to decrease with increasing pedestal radius, a result of the increased contact area through which energy may radiate. At the same time, for a contact areas with radius near 200~nm, the RBM mixes with a secondary ('pedestal') mode that displays a large vertical displacement at the disk center, as observed in the profile of mode \textcircled{\footnotesize{\hspace{0.008in}A}}. Such mixing is evidenced by an anti-crossing between the green and blue mechanical frequency curves in Fig. 1(e), and by a steep decrease in quality factor in the neighborhood of the anti-crossing. Reduced quality factors in this range are associated with the relatively large vertical displacement of the pedestal mode. For radii below 200~nm the RBM quality factor increases again (green curve in Fig. 1(e)), as the pedestal mode is driven towards lower frequencies.

\begin{figure*}[t!]
\centerline{\includegraphics[width=17 cm]{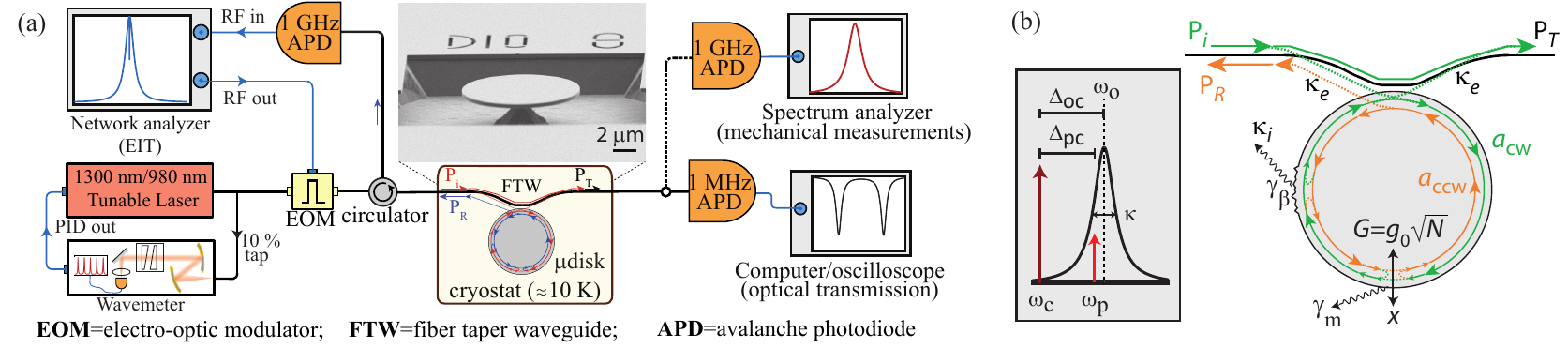}}
\caption{(a) Schematic of experimental setup for EIT measurements. (b) (Left) Sideband spectroscopy schematic. (Right) Schematic of the microdisk with the optical whispering gallery modes and optical and mechanical coupling mechanisms. $a_{\text{cw}}$/$a_{\text{ccw}}$ are amplitudes for clockwise/counterclockwise (cw/ccw) optical modes, which couple to the radial breathing mechanical mode (amplitude $x$) at a rate $g_{\text{0}}$ that is parametrically-enhanced to a rate $G$ through an optical control field (at $\omega_{\text{c}}$) that injects $N$ photons into the cavity. $\kappa_{i}$ and $\kappa_{e}$ are the intrinsic optical loss and waveguide-cavity optical coupling rates, respectively, $\gamma_{\text{m}}$ is the mechanical damping rate, and $\gamma_{\beta}$ is a backscattering rate that couples the cw and ccw optical modes.  $P_{i}$/$P_{R}$/$P_{T}$ are the incident, reflected, and transmitted optical powers through the FTW. }
\label{fig:SFig2}
\end{figure*}

While the geometrical details of the Si pedestal must be included for an exact determination of the pedestal mode frequencies, we have observed that, generally, smaller pedestal top radii were necessary to ensure that such modes would be out of the range of the RBM.  Our highest mechanical quality factors ($Q_m\approx1\times$10$^4$) were measured for a pedestal radius of 100~nm, which is essentially the smallest size that we can produce with our current fabrication process, where the theoretical value is $Q_m\approx4\times$10$^4$.

\section{Device fabrication}

The device fabrication started with a bare silicon wafer, as shown in Fig.~\ref{fig:SFig1}. A layer of 350~nm thick stoichiometric Si$_3$N$_4$ was grown by low pressure chemical vapor deposition (LPCVD), with a process-induced internal tensile stress of $\approx800$~MPa, as measured by the wafer bowing method. A 500~nm thick positive-tone electron beam (E-beam) resist was spin-coated on the Si$_3$N$_4$ film, followed by E-beam lithography and development in hexyl acetate at 8~$^{\circ}$C. The patterns were then transferred into the Si$_3$N$_4$ layer by an O$_2$/CHF$_3$/Ar inductively-coupled plasma reactive ion etch (RIE). This RIE step apparently leaves a thin layer of SiO$_2$ on top of the exposed Si surface, preventing any subsequent KOH undercut (Fig.~\ref{fig:SFig1}(b)). To rectify this, before the e-beam resist was removed, an additional SF$_6$/C$_4$F$_8$ inductively-coupled plasma RIE was carried out to remove the SiO$_2$ layer (as well as some Si), while the residual E-beam resist protected the Si$_3$N$_4$ device layer. After resist removal using a stabilized H$_{2}$SO$_{4}$/H$_{2}$O$_{2}$ solution, the sample was undercut in a 20~$\%$ KOH bath, as this concentration has an etch rate that is relatively insensitive to temperature fluctuations. The KOH etch was performed in two steps to achieve a small ($\lesssim200$~nm) top pedestal diameter under the microdisk. We started to etch the sample at 40~$^{\circ}$C to quickly remove the bulk Si, with periodic inspection of the pedestal size under an optical microscope every 6 to 10 minutes. When the pedestal size was less than 1~$\mu$m, the undercut was continued at a temperature of 20~$^{\circ}$C to reduce the etch rate. The undercut was stopped every 4 minutes for inspection, until one or more microdisks were completely released (etch rate variations across the chip prevent all devices from being completely released for the same undercut time). The pedestal diameters for the remaining (intact) microdisks were typically $\lesssim$200~nm, as determined by imaging with a scanning electron microscope.

\section{EIT experimental setup}

\begin{figure*}[t!]
\centerline{\includegraphics[width=15 cm]{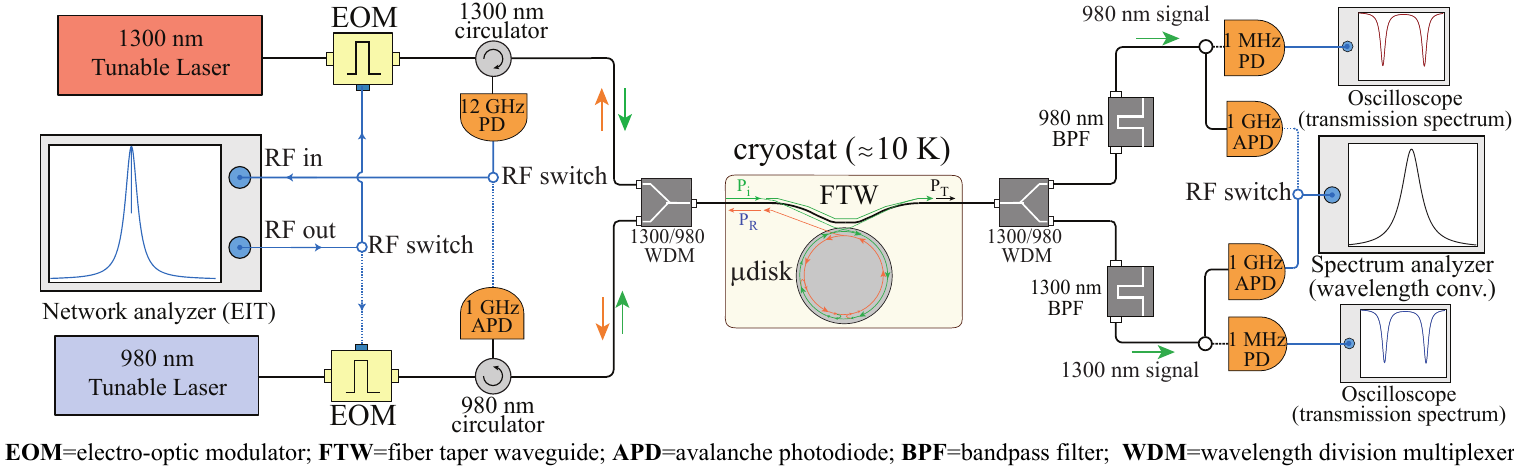}}
\caption{Detailed schematic of the wide-band wavelength conversion experiment. EIT spectroscopy in the 1300~nm and 980~nm bands is performed on the reflected signal from the cavity, with fast switching between the two bands enabled by radio frequency (RF) switches. In frequency upconversion (downconversion), the 1300~nm (980~nm) laser is modulated to generate an input probe signal field, which is converted to the 980~nm (1300~nm) band through application of the 980~nm (1300~nm) control field.  The frequency converted field transmitted past the cavity is measured on the RF spectrum analyzer.}
\label{fig:SFig3}
\end{figure*}

The experimental setup for device characterization and EIT measurements is shown in Fig.~\ref{fig:SFig2}(a). Light from a 1300~nm or 980~nm band tunable diode laser is coupled to the devices using an optical fiber taper waveguide (FTW). Optical modes are measured by sweeping the laser wavelength and recording the transmitted signal through the FTW. Mechanical modes are measured with the laser wavelength tuned to the shoulder of an optical mode. The transmitted optical power, which fluctuates due to the disk mechanical motion, is spectrally resolved on a real-time electronic spectrum analyzer. Increasing the optical power from $\approx0.2~\mu$W to higher powers when the laser is blue-detuned with respect to the cavity mode results in a narrowing of $Q_{\text{m}}$ by over an order of magnitude (Fig.~1(d)), as the laser drives the system into regenerative oscillations~\cite{ref:Kippenerbg_Vahala_Science}.  On the other hand, strong pumping on the red-detuned side of the cavity is limited by instability due to thermo-optic effects.  Mitigation of this thermo-optic instability is the primary reason why the EIT and wavelength conversion measurements presented in this paper are conducted with the sample housed in a liquid helium flow cryostat.

In EIT measurements, the probe is derived from the control field using an electro-optic amplitude modulator (EOM) to generate higher and lower frequency sidebands, only one of which is coupled into the cavity since it operates in the resolved sideband limit (Fig.~\ref{fig:SFig2}(b)). The EOM is driven by a network analyzer, with the modulation frequency swept to vary the probe-control beam detuning $\Delta_{\text{pc}}=\omega_{\text{p}}-\omega_{\text{c}}$. This results in sweeping the probe wavelength across the optical cavity mode when the the control wavelength is fixed. The probe and control fields pass through a circulator before going to the device, whose reflected signal is demodulated by the network analyzer to monitor the change of reflected probe signal during the sweep of the probe signal. A wavemeter tracks the control field wavelength and is used in feedback to lock $\Delta_{\text{oc}}$ (see section below).

\section{Locking of $\Delta_{\text{oc}}$}

In EIT experiments, the microdisk cavity is heated when the control laser reaches high power levels, resulting in a red-shift of the resonances due to the thermo-optic effect. It is thus difficult to maintain a fixed detuning ($\Delta_{\text{oc}}$) between the control field and optical cavity without feedback. To reduce this instability, we lock $\Delta_{\text{oc}}$ by adjusting the control laser wavelength to track the red-shifted cavity resonance, following a procedure similar to that outlined in ~\cite{ref:Chan_thesis}. First, a small portion of the control field is split off and fed to a wavemeter (Fig.~\ref{fig:SFig2}) equipped with a PID controller whose output is fed back to the laser to fix its absolute wavelength at a specified value.  This specified value is determined by sweeping $\Delta_{\text{pc}}$ over a broad frequency range (0 to 3 GHz) and fitting the acquired data by the vector network analyzer to determine the real value of $\Delta_{\text{oc}}$. The control field wavelength is then adjusted to ensure that the desired $\Delta_{\text{oc}}$ is achieved, and the process repeated until the fit value of $\Delta_{\text{oc}}$ remains stable over several scans in a row. Typical locking times are on the order of a few seconds (for lower powers) up to a couple tens of seconds (for the highest powers).  Power levels beyond that shown in Fig.~2 result in thermally unstable behavior, where the cavity mode shifts too far away from the control laser for the above approach to be effectively applied.

Once the locking procedure kept $\Delta_{\text{oc}}$ within a predetermined acceptable range around the target value (e.g., 50~MHz), EIT measurements consist of sweeps of $\Delta_{\text{pc}}$ recorded over two frequency ranges - the first over a broad range to provide an overall picture of the cavity reflection spectrum, and the second over a narrow range to record the details of the EIT dip. For each set of parameters (optical powers and $\Delta_{\text{oc}}$), these sweeps were repeated 16 times and the data plotted in Fig.~2 is an average of these scans.

\section{wavelength conversion experimental setup}

In frequency upconversion (downconversion), a 1285~nm (990~nm) control pump laser is modulated to generate an input probe signal field in a similar fashion to the EIT experiments. The input probe is then upconverted (downconverted) through application of a 990~nm (1285~nm) control pump field. The detuning between input probe field and 1285~nm (990~nm) control pump is swept to assess the bandwidth of the conversion process.

The detailed setup used for wavelength conversion experiments is shown in Fig.~\ref{fig:SFig3}. The following discussion is based on the upconversion process, while a similar explanation can be readily applied to downconversion by swapping the input signal and converted signal wavelengths. As in Fig.~\ref{fig:SFig2}, EIT spectroscopy in the 1300~nm and 980~nm bands is performed on the reflected signal from the cavity. Fast switching between the two bands is enabled by RF switches, and the main results of the EIT measurements are to determine $\Delta_{\text{oc}}$ and the cooperativity achieved for each mode. Wavelength conversion experiments proceed by combining the input 1285~nm probe field with the control fields in both the 980~nm and 1300~nm bands using a wavelength division multiplexer (WDM) before being sent into the device. Light exiting the optomechanical system is spectrally separated into the 980~nm and 1300~nm bands using a WDM and bandpass filters (BPFs), and the converted 990~nm signal is detected by a 1~GHz APD whose output is recorded by a real-time electronic spectrum analyzer. The input probe wavelength (near 1285 nm) is swept while the control wavelengths in both bands are fixed, which results in a sweep of the generated probe signal at the conversion wavelength (990~nm). The transmitted converted signal at 990~nm is measured by the spectrum analyzer. This measured signal is the result of interference between the 990~nm pump, which is constant in both power and wavelength, and the converted 990~nm probe. As sweeping of the 1285~nm input probe signal results in a sweep of the converted 990~nm tone, we are able to assess the dependence of the conversion efficiency on the input signal-control detuning ($\Delta_{\text{pc}}$) at 1285 nm. Each curve shown in Fig.~3(c) is the envelope of $\approx$~10 swept traces of the measured signal at 990 nm.

\section{wavelength conversion measurements}

In wavelength conversion experiments, the RF spectrum analyzer measures a signal that results from the interference of the converted probe signal with the control pump situated in the same wavelength band (and which is detuned by a mechanical frequency $\omega_{\text{m}}$).  Considering the case of upconversion from 1285~nm to 990~nm, the optical power measured by the AC-coupled APD is proportional to $\sqrt{P_{\text{990,conv}}P_{990}}$, where $P_{\text{990,conv}}$ is the wavelength-converted probe signal and $P_{990}$ is the control pump power.  The APD detector converts this to a voltage, and the RF spectrum analyzer measures an RF power proportional to the square of this voltage, that is, proportional to $P_{\text{990,conv}}P_{990}$.  Next, we can write $P_{\text{990,conv}}=\eta_{\text{up}}P_{\text{1285,input}}\omega_{990}/\omega_{1285}$, where $P_{\text{1285,input}}$ is the input probe signal power, $\omega_{990}$ ($\omega_{1285}$) is the optical frequency at 990~nm (1285~nm), and $\eta_{\text{up}}$ is the photon number conversion efficiency (the ratio of the converted signal photon number to the input signal photon number).  Finally, because $P_{\text{1285,input}}$ is created through modulation of the 1285~nm control field, the two are related by the modulation index $\beta_{1285}$ as $P_{\text{1285,input}}=\beta_{1285}^2P_{1285}$, where $P_{1285}$ is the 1285~nm control field power.

Putting this all together, we write the measured RF power at $\omega_{\text{m}}$ as:

\vspace{-0.2in}
\begin{equation}
H_{990}=\Re_{990}\eta_{\text{up}}\beta_{1285}^2P_{1285}P_{990}\frac{\omega_{990}}{\omega_{1285}} \nonumber
\end{equation}
\vspace{-0.15in}

\noindent $\Re_{990}$ is a constant that depends on the APD gain, responsivity, and load resistance.

A similar expression can be written in the case of frequency downconversion, swapping the roles of the 990~nm and 1285~nm bands, and using a prefactor $\Re_{1285}$ that depends on the detector response at 1285~nm.  The efficiencies of upconversion and downconversion, $\eta_{\text{up}}$ and $\eta_{\text{down}}$, can thus be determined from the RF spectra, control pump powers, optical frequencies, and input modulation indices.

\section{wavelength conversion data}

\begin{figure}
\centerline{\includegraphics[width=8 cm]{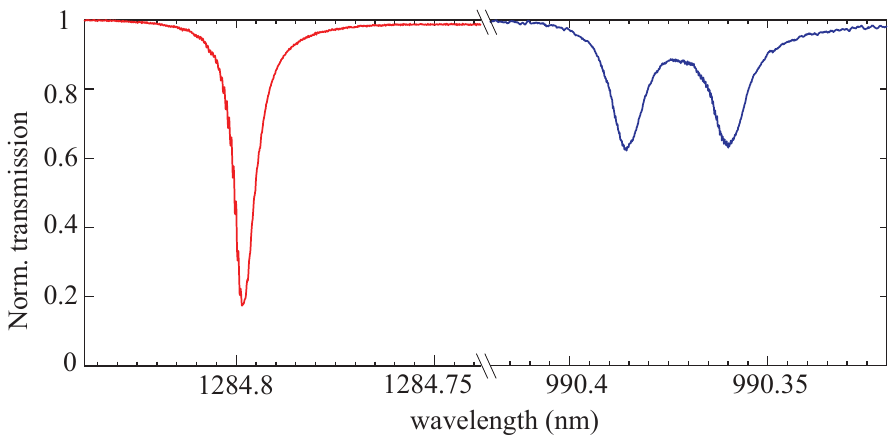}}
\caption{Measured optical spectra of the two modes, one at 990~nm and the other at 1285~nm, in the wavelength conversion experiment.}
\label{fig:SFig4}
\end{figure}

The measured optical spectra of the two modes used in the wavelength conversion experiments is shown in Fig.~\ref{fig:SFig4}.  In comparison to Fig.~1 and 2, $Q_{\text{o}}$ here is lower, with $\kappa/2\pi\approx1$~GHz and 3~GHz at 1285~nm and 990~nm, respectively. The increased losses are both due to additional sample degradation and the need to simultaneously achieve a reasonable coupling level to both optical modes (in Fig.~2, this was individually optimized).

\begin{figure*}
\centerline{\includegraphics[width=15 cm]{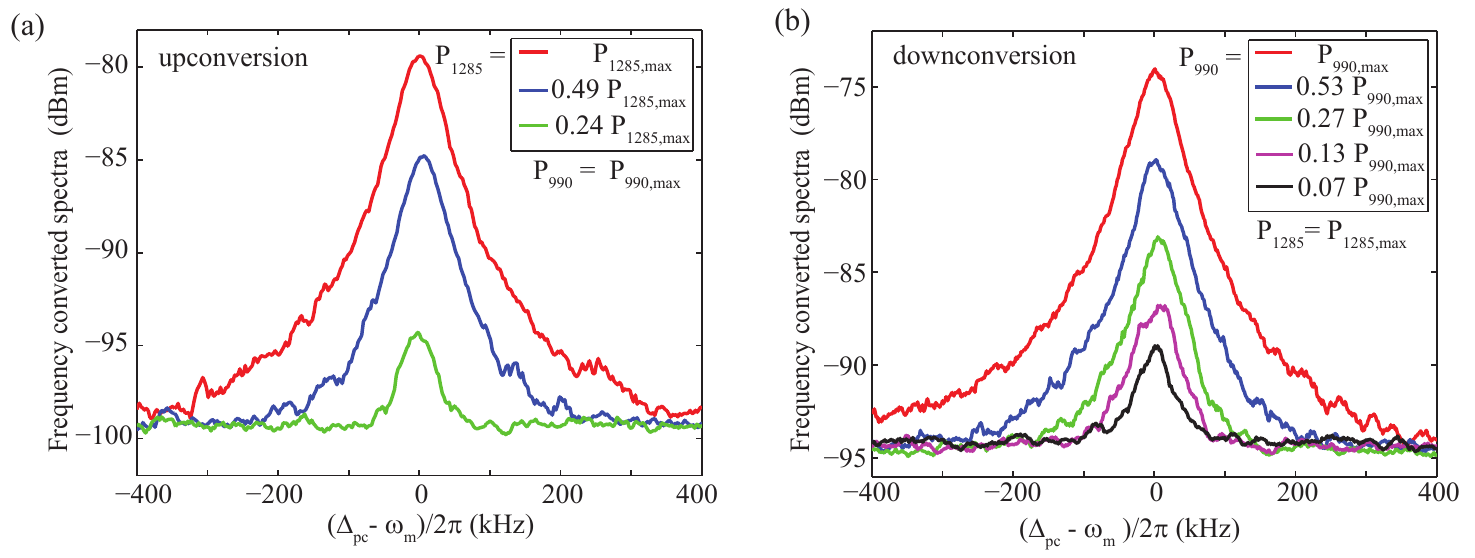}}
\caption{(a) Upconversion signals (990 nm) measured at different optical powers of the 1285~nm control field at the device input. (b) Downconversion signals (1285 nm) measured at different optical powers of the 990~nm control field. In upconversion (downconversion), the 990~nm (1285~nm) control field power is fixed at $P_{\text{990,max}}$=9.0~$\mu$W ($P_{\text{1285,max}}$=63.8~$\mu$W ) and the 1285~nm (990~nm) control field power is stepped.}
\label{fig:SFig5}
\end{figure*}

Wavelength conversion measurements were taken as a function of detuning between the input probe and control field within the same wavelength band ($\Delta_{\text{pc}}$) and as a function of pump power for both control fields.  Figure~3(c) shows the dependence of the converted signal on the control power in the converted wavelength band, while here we show the conversion dependence on the control power in the input signal band (Fig.~\ref{fig:SFig5}). The acquired spectra are the result of interference of the converted probe signal and control pump in the conversion band, for different power levels of the control pump situated in the input signal band. Therefore, different from the results shown in Fig.~3(c), all the curves in each figure of Fig.~\ref{fig:SFig5} share the same noise floor level because the control pump power in the converted band is fixed.

\section{Fitting the data}

Experimentally-measured optical transmission spectra are fit using coupled mode theory for a resonator-waveguide system.  For single dips, this yields a Lorentzian function, while doublet modes are fit using a model that takes into account backscattering due to surface roughness, as described elsewhere~\cite{ref:Borselli,ref:Kippenberg}.  In the limit of large doublet spitting with respect to the cavity linewidths, a pair of Lorentzians can accurately fit the data.  Mechanical spectra were fit with single Lorentzian functions.

Fitting of the EIT signals (Fig. 2(b)) is based on Eq. (1) from the main text, which we repeat here:

\begin{equation}
r(\Delta_{\text{pc}}) = -\frac{1}{1+\frac{2i(\Delta_{\text{oc}}-\Delta_{\text{pc}})}{\kappa}+\frac{C}{\frac{2i(\omega_{\text{m}}-\Delta_{\text{pc}})}{\gamma_{\text{m}}}+1}}
\label{eq:Seq1}
\end{equation}

\noindent The values of $\kappa$, $\Delta_{\text{pc}}$, and $\Delta_{\text{oc}}$ were acquired from the broad scans of the EIT signal. The intracavity photon number is determined as:

\begin{equation}
N = \frac{1}{\hbar\omega_{o}}\sqrt{\xi}\Delta TQ_{i}(\frac{P_{\text{in}}}{\omega_o})\frac{1}{1+(\frac{\Delta_{\text{oc}}}{2\kappa})^2}
\label{eq:Seq2}
\end{equation}

\noindent where $\hbar$ is the Planck constant divided by 2$\pi$, $\xi$ is the FTW transmission, $\Delta T$ is the depth of the optical resonance in the transmission spectrum ($\Delta T$=1 at critical coupling), $Q_i$ the intrinsic optical $Q$, and $P_{\text{in}}$ is the optical power at the FTW input. We measured a background signal with the FTW far away from the device to account for the frequency-dependent response of the APD, EOM, and network analyzer. The fitting was carried out on the experimentally acquired EIT data divided by the background signal.

The fitting of the EIT width versus control power (inset in Fig. 2(c)) is based on the expression for the total mechanical damping rate $\gamma=\gamma_{\text{m}}(1+C)$. The fitted value of $g_{\text{0}}$ from the EIT width is approximate equal to that obtained from the EIT depth fitting, with an error of $\approx6~\%$.

\section{Uncertainty estimates}

\begin{figure}[b!]
\centerline{\includegraphics[width=8.5 cm]{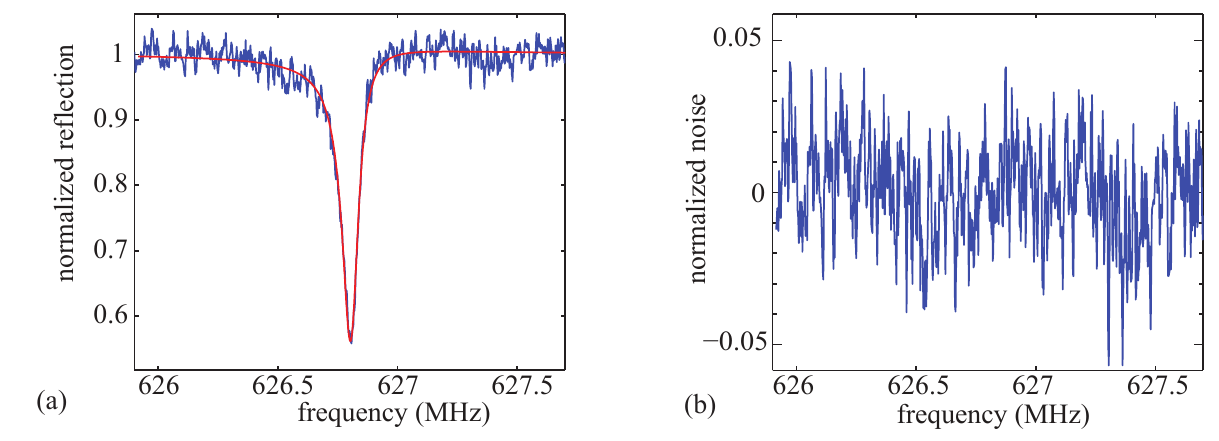}}
\caption{(a) Raw data (blue) and the fitted curve (red) of the 1300 nm EIT signal at the control power of 21.9~$\mu$W. (b) Difference between the raw data and the fitted curve shown in (a).}
\label{fig:SFig6}
\end{figure}

The error bars in Fig. 2(c) arise from the noise of recorded raw EIT data. By subtracting the fitted curve from the raw EIT data (Fig.~\ref{fig:SFig6}(a)), we obtained the noise of the EIT signal (Fig.~\ref{fig:SFig6}(b)). The one standard deviation value of this noise signal is considered as the uncertainty of the EIT signal depth and is plotted as the error bar in Fig. 2(c). By combining the uncertainty of the reflected EIT signal and the slopes of the fitted curves at both the left and right shoulders, we obtained the uncertainty of the width of the EIT signal. This width uncertainty is plotted as the error bars in the inset figure of EIT width versus the optical power in Fig. 2(c).

The uncertainty in Q$_{\text{m}}$ shown in Fig. 1(d) is given by the 95~$\%$ confidence intervals of the Lorentzian curve fitting of the experimentally measured mechanical spectra.

\end{document}